\documentclass[conference]{IEEEtran}
\IEEEoverridecommandlockouts
\usepackage{cite}
\usepackage{amsmath,amssymb,amsfonts}
\usepackage{algorithmic}
\usepackage{graphicx}
\usepackage{textcomp}
\usepackage{booktabs}
\usepackage{xcolor}
\def\BibTeX{{\rm B\kern-.05em{\sc i\kern-.025em b}\kern-.08em
    T\kern-.1667em\lower.7ex\hbox{E}\kern-.125emX}}
\begin{document}

\title{On the Sensitivity of Deep  Load Disaggregation  to Adversarial Attacks\\
}

\author{\IEEEauthorblockN{Hafsa Bousbiat}
\IEEEauthorblockA{
\textit{University of Klagenfurt}\\
Klagenfurt, Austria \\
hafsa.bousbiat@edu.aau.at}
\and
\IEEEauthorblockN{Yassine Himeur}
\IEEEauthorblockA{
\textit{University of Dubai}\\
Dubai, UAE \\
yhimeur@ud.ae}
\and
\IEEEauthorblockN{Abbes Amira}
\IEEEauthorblockA{
\textit{University of Sharjah}\\
Sharjah, UAE \\
aamira@sharjah.ac.ae}
\and
\IEEEauthorblockN{Wathiq Mansoor}
\IEEEauthorblockA{
\textit{University of Dubai}\\
Dubai, UAE \\
wmansoor@ud.ac.ae}
}

\maketitle

\begin{abstract}
Non-intrusive Load Monitoring (NILM) algorithms, commonly referred to as load disaggregation algorithms, are fundamental approaches for effective energy management. Despite the success of deep models in load disaggregation, they face various challenges, particularly those pertaining to privacy and security. This paper investigates the sensitivity of prominent deep NILM baselines to adversarial attacks, which have proven to be a significant threat in domains such as computer vision and speech recognition. 
Adversarial attacks entail the introduction of imperceptible noise into the input data with the aim of misleading the neural network into generating erroneous outputs. We investigate the Fast Gradient Sign Method (FGSM), a well-known adversarial attack, to perturb the input sequences fed into two commonly employed CNN-based NILM baselines: the Sequence-to-Sequence (S2S) and Sequence-to-Point (S2P) models. Our findings provide compelling evidence for the vulnerability of these models, particularly the S2P model which exhibits an average decline of 20\% in the F1-score even with small amounts of noise. Such weakness has the potential to generate profound implications for energy management systems in residential and industrial sectors reliant on NILM models.
\end{abstract}


\section{INTRODUCTION}

Providing appliance-level feedback has been widely acknowledged as a promising approach for achieving significant energy savings~\cite{himeur2022techno}. The previous strategy can lead up to 12\% of energy savings in the residential sector \cite{himeur2022recent, gopinath2020energy}. It allows individuals to make informed decisions about their daily interaction with in-home electrical devices. In this regard, Non-intrusive Load Monitoring (NILM) has emerged as a promising technology that utilizes advanced algorithms to infer the power consumption of individual appliances based on aggregate household measurements~\cite{bousbiat2023neural, bousbiat2022neural}. With  minimal hardware requirements, NILM approaches are an attractive solution for energy management and sustainability efforts. More precisely, NILM algorithms rely only on a single metering point reporting the total power consumption of the household to identify operating appliances \cite{bousbiat2022neural} allowing thus to avoid complex hardware deployment and maintance.

Exceptional performance enhancement in NILM scholarship can be observed in recent years. Particularly, the adoption of deep models for load disaggregation by Kelly et al. \cite{kelly2015neural} in 2015 has been a turning point~\cite{ bousbiat2023neural, himeur2022recent}. Compared to statistical classical models, deep neural networks have shown tremendous potential in enhancing the accuracy of energy consumption estimates~\cite{batra2019towards}. A main advantage of these models is the automatic feature extraction that provides an end-to-end solution. 
Nonetheless, the robustness of these models against adversarial attacks has become a growing concern in recent years \cite{tan2022adversarial}.
Adversarial attacks can pose a significant threat to NILM's accuracy and reliability by injecting small amounts of noise into the input data to deceive the neural network and generate erroneous outputs~\cite{singh2021adversarial}.  In the case of NILM, adversarial attacks can cause wrong energy consumption estimates~\cite{ou2020singular}, privacy breaches, instability in home energy management systems~\cite{kotak2022adversarial, kaselimi2022towards}, and even safety risks~\cite{liaqat2022event} when used in critical infrastructures.
 This paper aims to address this gap by evaluating the sensitivity of two popular NILM baselines to adversarial attacks. 

The remainder of this paper is organized as follows: Section \ref{sec:related} discusses related work and recent literature providing insights into the state-of-the-art  NILM models and adversarial attacks. Section \ref{sec:methodology} describes the different steps followed to perform the attack, including the generation of adversarial examples and their evaluation on the target models.  Section \ref{sec:evaluation} describes the experimental setup followed by a presentation the results of the experimental setup in Section \ref{sec:results}. Finally,  Section \ref{sec:conclusion} concludes the paper by summarizing the main contributions of this research and discussing potential future directions in this area.

\section{RELATED WORK}
\label{sec:related}

\begin{figure*}
    \centering
    \includegraphics[scale=.65]{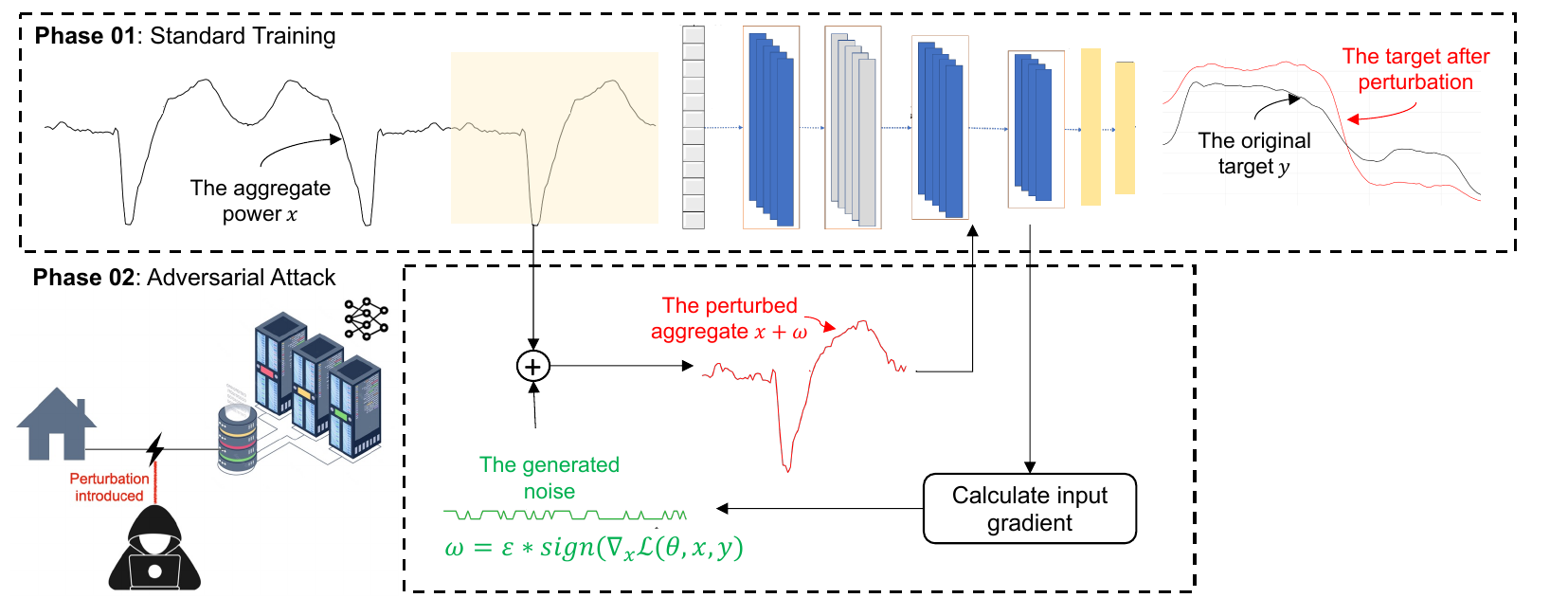}
    \caption{Flow chart of the performed attack}
    \label{fig:methodology}
\end{figure*}

Non-intrusive load monitoring (NILM) gained significant attention in the past five years, with neural networks being widely acknowledged as state-of-the-art models \cite{huber2021review, bousbiat2023neural}. Popular deep baselines for NILM, such as Sequence-to-Sequence and Sequence-to-point, are based on convolutional layers \cite{kelly2015neural}. However, other contributions based on LSTM \cite{kaselimi2020context} have also demonstrated the robustness of these layers in identifying appliance usage. Advanced models, such as the BERT model \cite{yue2020bert4nilm} and UNET model \cite{faustine2020unet}, have also been suggested in the literature.

Despite the success of these models in estimating the power consumption of household appliances under different conditions \cite{batra2019towards}, they suffer from privacy and security issues \cite{bousbiat2022ageing}. They can reveal sensitive information about individuals' daily routines, as illustrated in \cite{schirmer2021identification, bousbiat2022ageing, bousbiat2020augmenting}, leading to concerns from the consumer's side. 
The security of NILM systems is thus a main concern when deep models are adopted. Yet, it received little attention. A compact review of adversarial attacks in smart grid scenarios was suggested in \cite{hao2022adversarial}, considering deep neural networks. The review study revealed that despite the potential for extreme harm, these attacks have received very little attention from the research community. Two types of attacks can be performed: white-box and black-box attacks.

White-box attacks assume full knowledge of the attacked model \cite{akhtar2018threat}, and are gaining interest in related work. They were particularly considered in \cite{singh2021adversarial}, targeting IoT devices where the case of residential smart meters was considered as a case study. The obtained results demonstrated that adversarial samples are often indistinguishable from real samples, leading to high success rates of the performed attack. This finding illustrates that adversarial attacks represent a major threat for energy management systems (EMS) both in residential and industrial setups, notably in the case of sensitive sectors, as they can easily be used to fool demand response programs. However, this attack model makes the assumption of complete knowledge of the model, which is hard to obtain in real scenarios.
In contrast, black-box attacks assume no knowledge of the model or the training data \cite{akhtar2018threat}. The main advantage of adopting this threat model, compared to the previous one, is its practicality. The significant influence of this attack on energy analytics and load disaggregation was explored in \cite{wang2021stealthy, wang2020deep}. The findings revealed that black-box attacks can be performed even in cases where the adversary has limited knowledge of the target system.


\section{METHODOLOGY}
\label{sec:methodology}

The  threat model aims to fool neural networks into producing inaccurate estimates of power consumption, thereby disrupting the energy management systems of smart buildings. 
As the attack involves adding perturbations to the input data, the attacker's primary challenge would be to determine the optimal amount of noise to introduce, maximizing the success rate while remaining undetected by data poisoning detection methods. The current paper's methodological design aims to evaluate this aspect while also assessing whether the attack's effect is similar for two different disaggregation models, given appliances with heterogeneous power consumption magnitudes. 

The steps involved in the methodology are depicted in Figure \ref{fig:methodology}. Firstly, the models are trained on the original data and evaluated on a testing data set, without attack. Secondly, the gradient of the pre-trained model is used to generate perturbed inputs, which are then fed back into the network to produce the corresponding predictions.

\subsection{Models}
The Seq2Point and Seq2Seq \cite{kelly2015neural} models for NILM  have gained significant attention due to their competitive performance in estimating the power consumption of household appliances. These models are based on convolutional layers and have similar structures, differing only in the shape of their output. 
The choice of the Seq2Point and Seq2Seq models as representative models for load disaggregation is a solid choice, given their competitive performance, wide acknowledgement, and adoption in the research community. To implement these models, we  leveraged the code available from NILMtk \cite{bousbiat2022unlocking}, a popular open-source toolkit for NILM research. The training process was conducted for 150 epochs, using an Adam optimizer with a learning rate of $10^{-4}$.

\subsection{Attack}

We propose to assess the robustness of the baseline models against a widely recognized white-box adversarial attack, namely the Fast Sign Gradient Sign Method (FGSM). This attack leverages the gradient of the model to generate an adversarial sample of the input. For time series data, this technique involves computing the gradient of the loss function $\mathcal{L}$ with respect to the input sequence, in order to create a new sequence that increases the loss. This newly generated sequence is referred to as an adversarial sample $\Tilde{x}$, where $\epsilon$ is a multiplier to ensure that the added perturbations remain small.

\begin{equation}
    \Tilde{x} = x + \epsilon * sign (\nabla_x \mathcal{L}(\theta, x, y))
\end{equation}

The FGSM has an important feature, namely that the gradient calculation is performed with respect to each input sequence, enabling the attacker to identify how each value of the input sequence contributes to the loss function, with the ultimate goal of misleading the model.
It is worth noting that the FGSM aims to mislead an already pre-trained model, without affecting its parameters. Hence, it represents a powerful tool to assess the model's robustness against adversarial attacks, as it provides insight into the model's vulnerabilities without affecting its training process. 




\section{Evaluation}
\label{sec:evaluation}
\subsection{Data}

The present study utilizes data from UKDALE, a dataset recorded in the UK, comprising five buildings, with the first building having a recording period longer than three years. The pre-trained models are constructed using data from the first building, where a four-month recording period with a sampling rate of 30 seconds is considered. The study further considers four appliances with varying consumption patterns: the washing machine, the kettle, the microwave, and the fridge. For testing purposes, a period of one month is considered for each appliance. 


    

\subsection{Metrics}

For a fair evaluation of the models as well as the effect of the FGSM attack, three metrics are used,  the Mean Average Error (MAE), the F1-score and the Normalised Disaggregation Error (NDE).
The MAE for an appliance $i$ is described as follows:

\begin{equation}\label{eq:mae}
    \text{MAE}^{(i)} = \frac{1}{N} \cdot \sum_{t=0}^{N-1} |\hat{y}_t^{(i)}-y_t^{(i)}| 
\end{equation}

where, $y_t$ is the ground truth values of the power consumption, $\hat{y}_t$ is the predicted power consumption, and $N$ represents the number of samples. In addition to the MAE, the NDE is considered as recommended in recent literature \cite{klemenjak2020towards}:

\begin{equation}
\text{NDE}^{(i)} =  \sqrt{\frac{\sum_{t=0}^{N-1}{(\hat{y}_t^{(i)}-y_t^{(i)}})^2}{\sum_{t=0}^{N-1}{(y_t^{(i)}})^2}}
  \label{equa:f1}
\end{equation}

Furthermore, the F1-score is used to assess the performance of the model on state estimation of different appliances, defined as follows:

\begin{equation}
    F1-score = \frac{2 \cdot Precision \cdot Recall}{Precision + Recall} 
\end{equation}

Where the Precision = TP/(TP+FP), Recall = TP/(TP+FN). A threshold of 10 \emph{watts} was used to derive states of the fridge and a threshold of 500 \emph{watts} was used in the case of the remaining appliances.

\begin{figure}
    \centering
    \includegraphics[scale=.38]{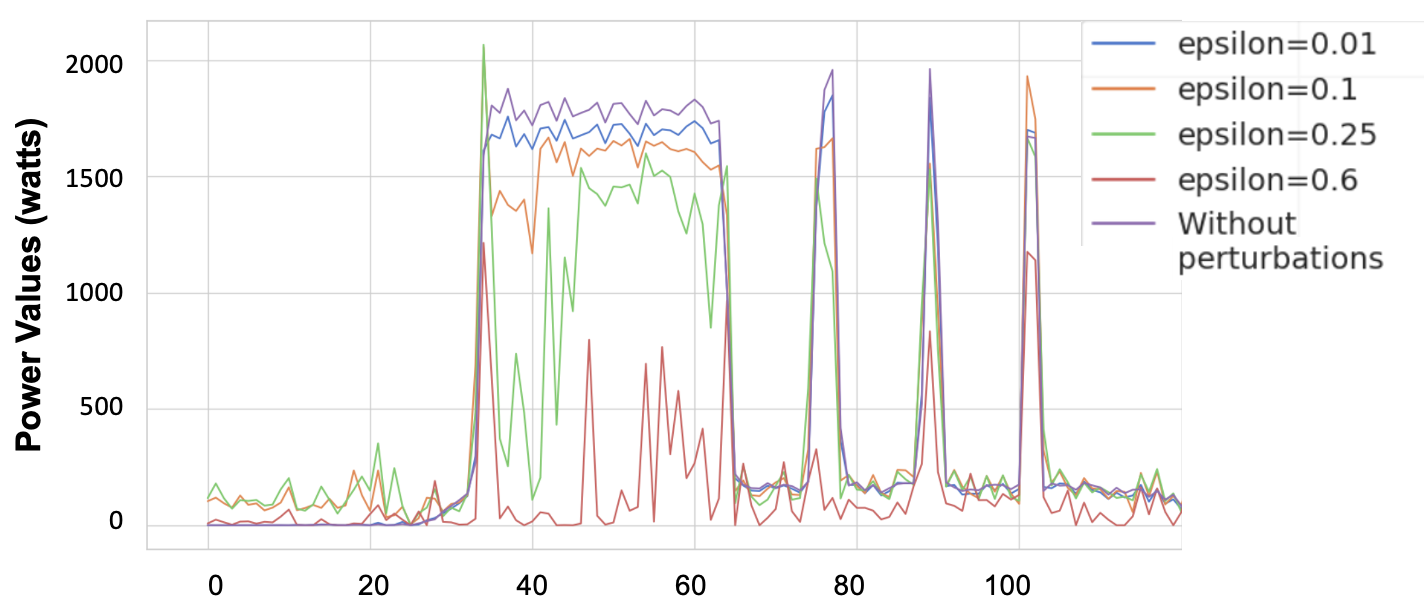}
    \caption{Seq2Point model  for the washing machine}
    \label{fig:seq2point}
\end{figure}
\begin{figure}
    \centering
    \includegraphics[scale=.38]{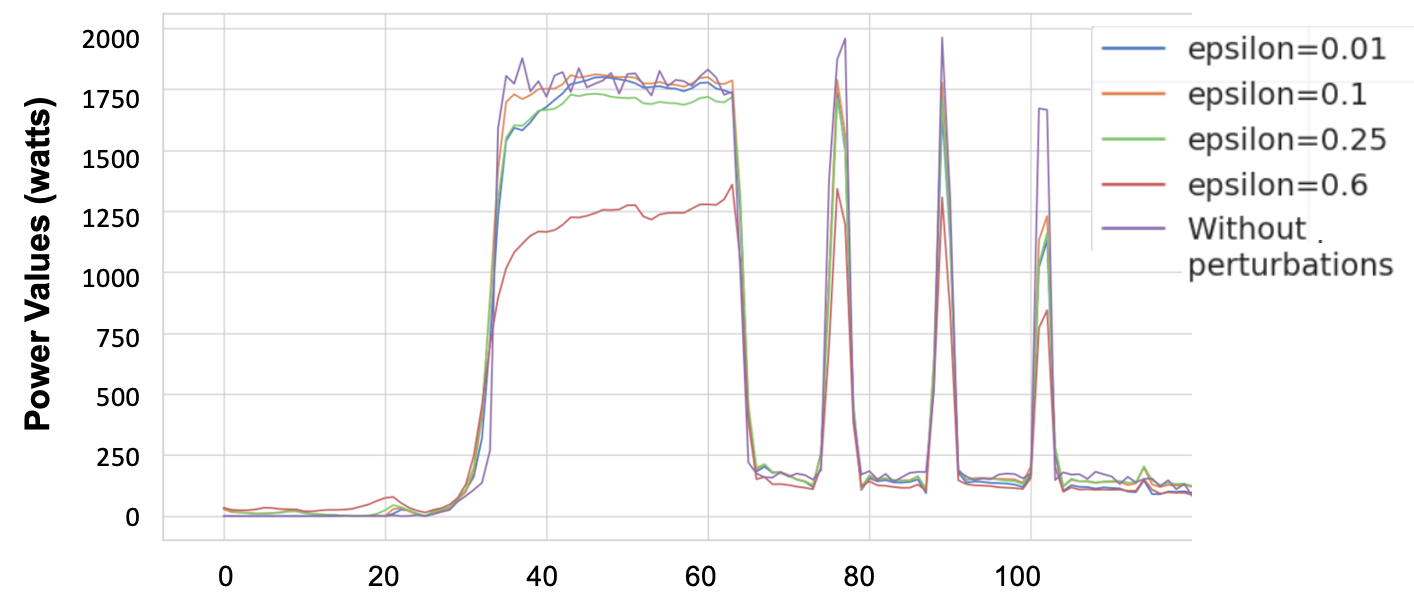}
    \caption{Seq2Seq model for the washing machine}
    \label{fig:seq2seq}
\end{figure}

\section{RESULTS \& DISCUSSION}
\label{sec:results}

The results obtained for both baseline models are presented in Table \ref{tab:results}. We report on four scenarios: the performance of the model without attack and the performance of the model under attack with three different values of $\epsilon$ (0.01, 0.10, and 0.25). For each of the considered scenarios, we report on the three metrics for the four appliances considered.

\begin{table*}[ht]
\caption{Obtained results for Seq2Point models}
    \label{tab:results}
    \centering
    \begin{tabular}{lcccccccccccccccc}
    \toprule&&  \multicolumn{3}{c}{Without Attack}  & & \multicolumn{3}{c}{FGSM ($\epsilon=0.01$)} & &  \multicolumn{3}{c}{FGSM ($\epsilon=0.10$)} & & \multicolumn{3}{c}{FGSM ($\epsilon=0.25$)}\\ \cmidrule{3-5} \cmidrule{7-9} \cmidrule{11-13} \cmidrule{15-17}
         && MAE& F1-score & NDE && MAE& F1-score & NDE && MAE& F1-score & NDE && MAE& F1-score & NDE \\ \hline \hline
        \multicolumn{17}{c}{\textbf{Seq2Point}} \\ \hline \hline
         Washing Machine && 12.4 & 0.67 & 0.73 && 14.20 & 0.63 & 0.75 && 126.63 & 0.39 & 1.17 && 153.94 & 0.13 & 1.40 \\ \hline
         Kettle && 05.4 & 0.86 & 0.38 &&  5.5 & 0.86 & 0.39 && 9.48 & 0.77 & 0.52 && 15.38 & 0.66 & 0.71\\ \hline
         Microwave && 06.0 & 0.63 & 0.71 && 7.57 & 0.60 & 0.79 && 11.68 & 0.34 & 1.02 && 12.41 & 0.22 & 1.04\\ \hline
         Fridge && 15.0 & 0.83 & 0.42 && 26.4 & 0.74 & 0.61 && 39.3 & 0.61 & 0.79 && 43.0 & 0.45 & 0.93\\ \hline \hline
         \multicolumn{17}{c}{\textbf{Seq2Seq}} \\ \hline \hline
         Washing Machine && 12.9 & 0.66 & 0.76 && 13.24 & 0.66 & 0.75 && 51.55 & 0.66 & 0.75 && 71.39 & 0.59 & 0.82\\ \hline
         Kettle          && 06.3 & 0.89 & 0.35 && 6.43 & 0.89 & 0.35 && 7.24 & 0.88 & 0.36 && 9.17 & 0.83 & 0.40\\ \hline
         Microwave       && 07.4 & 0.60 & 0.70 && 7.83 & 0.63 & 0.69 && 9.98 & 0.65 & 0.75 && 12.48 & 0.63 & 0.85 \\ \hline
         Fridge          && 17.6 & 0.79 & 0.43 && 17.33 & 0.77 & 0.43 && 29.59 & 0.62 &0.55 && 36.22 &0.60 & 0.63 \\ 
         
         \bottomrule
    \end{tabular}
    
\end{table*}

When no attack is performed,  both Seq2Point and Seq2Seq models yield equivalent performance considering the three metrics with the Seq2Point model providing slightly better results considering all appliances and all metrics. The latter observation is reflected by a minimal f1-score of 67\% and a maximum of 12.4 \emph{watts} for the MAE. The previous observation aligns well with existing literature \cite{batra2019towards} suggesting the superiority of Seq2Point model.

The results show that as the epsilon value increases, the performance of all appliances deteriorates with the Kettle performing the best and the Fridge performing the worst, for both models. At an epsilon value of 0.01, all appliances show a slight decrease in performance. Nonetheless, starting from an epsilon value of 0.10, the MAE of all appliances increases significantly, with the Washing Machine appliance showing the highest increase. Meanwhile, F1-score decreases for all appliances, with the Microwave appliance showing the highest decrease. At an epsilon value of 0.25, the MAE, F1-score, and NDE values continue to deteriorate for all appliances, with the Washing Machine appliance showing the highest MAE and the Fridge appliance showing the highest NDE.

When comparing the effect of the added noise on the Seq2Seq and Seq2Point models, the Seq2Seq shows more robustness against the attack. On one side, Seq2Point demonstrates higher sensitivity to the added noise with a decrease in all metrics by a factor of three for frequently used appliances (i.e., the fridge, the microwave, the kettle) and a decrease by factor of ten (10) in the case of the washing machine for $\epsilon=0.10$, considered as one of the major loads. On the other side, the Seq2Seq demonstrates moderate sensitivity where a decrease with only a factor of 1.5 is recorded for frequently used appliances and a factor of 5 in the case of the washing machine for the same values of $\epsilon$. It can be seen that even if high values of $\epsilon$ lead to slight deterioration in the MAE, the seq2seq model still yields approximately the same F1-score values unlike the Seq2Point model where the deterioration is recorded in all metrics. 

The Figure \ref{fig:seq2point} illustrates one generated activation for the washing machine for the Seq2Point model. It can be observed that the noise added first reflects on the OFF states through small fluctuations in the order of hundreds of \emph{watts} starting from values of $\epsilon$=0.10. Furthermore, the figure shows that higher values of $\epsilon$ lead to an attenuated signal or undetected states (e.g., second peak in the figure). Figure \ref{fig:seq2seq} illustrates the same activation generated with Seq2Seq model. It is clear that the model shows more robustness where it is capable of generating stable signal even with higher values of added noise with slight deterioration in the estimated values.

The adversarial attack illustrated in this study on NILM models can easily be leveraged in smart homes to mislead the neural networks. This is particularly relevant in the case of  home automation functions \cite{9128406}. For example, an attack can be performed on light control systems that are based on NILM and leads to considerable losses if unnoticed.
It can also lead to the incorrect estimation of load flexibility in the system. If a device is identified as "off" although it is actually "on," it will not be considered for automated control, which will affect the ability  to shave demand peaks when a large number of flexible loads are attacked. For instance, if the system estimates that 75 kW of demand during peak hours in a day stem from the heating system in winter time and the NILM algorithm misclassifies the state of this appliance, then the peak reduction potential is lost. Moreover, the attack can be leveraged at specific times to mislead occupancy detection systems and easily lead to erroneous predictions that can be used by malicious parties.

\section{CONCLUSION}
\label{sec:conclusion}
This paper suggested to assess the robustness of the Seq2Seq and Seq2Point models against adversarial attacks considering the FGSM attack. Even with smaller values of $\epsilon$, the obtained results reveal high sensitivity of the Seq2Point model, while demonstrating a moderate sensitivity of the Seq2Seq model. This study confirms findings from related about the superiority of Seq2Point in normal scenarios but highlights that Seq2Seq reveal more strength to injected noise. Adversarial attacks can have destructive effect on EMS systems and different energy services. In future work, we aim reinforce the robustness of these models to such attacks using different strategies including augmenting the set of training data with adversarial samples.

\bibliographystyle{IEEEtran}
\bibliography{references.bib}

\end{document}